\documentclass[article,aps,twocolumn,float,prd, %
nofootinbib]{revtex4}
\usepackage{tikz,color,cancel,acronym}
\usepackage[colorlinks,linkcolor=blue,citecolor=blue,urlcolor=blue ]{hyperref}
\usepackage{ulem} % used solely for the purpose of enable strike-through lines. Should remove/comment before submission
\usepackage{amsmath}
\allowdisplaybreaks[4] % float equations
\usepackage{graphicx}% Include figure files
\usepackage[capitalize]{cleveref}
\usepackage{booktabs}
\newcommand{\beq}{\begin{equation}}
\newcommand{\beqa}{\begin{eqnarray}}
\newcommand{\eeq}{\end{equation}}
\newcommand{\eeqa}{\end{eqnarray}}
\newcommand{\nn}{\nonumber}

\newcommand{\TRC}{MOE Key Laboratory of TianQin Mission, TianQin Research Center for Gravitational Physics $\&$ School of Physics and Astronomy, Frontiers Science Center for TianQin, Gravitational Wave Research Center of CNSA, Sun Yat-sen University (Zhuhai Campus), Zhuhai 519082, China}
\begin{document}

%\baselineskip 8mm
%%%%%%%%%%%%%%%%%%%%%%%%%%%%%%%%%%%%%%%%%%%%%%%%%%%%%%%%%%%%%%%%%%%%%%%%%%
%%%%%%%%%%%%%%%%%%%%%%%%%%%%%%%%%%%%%%%%%%%%%%%%%%%%%%%%%%%%%%%%%%%%%%%%%%

%\title{Assessing the detectability of the stochastic gravitational wave background \\
%with the TianQin I + II network}

\title{Detecting the stochastic gravitational wave background \\
with the TianQin-LISA detector network}
%detector. II. Multiple constellation detection}

%%%%%%%%%%%%%%%%%%%%%%%%%%%%%%%%%%%%%%%%%%%%%%%%%%%%%%%%%%%%%%%%%%%%%%%%%%
%%%%%%%%%%%%%%%%%%%%%%%%%%%%%%%%%%%%%%%%%%%%%%%%%%%%%%%%%%%%%%%%%%%%%%%%%%
%
%
%
%%%%%%%%%%%%%%%%%%%%%%%%%%%%%%%%%%%%%%%%%%%%%%%%%%%%%%%%%%%%%%%%%%%%%%%%%%
\author{Jun Cheng }
%\email
\thanks{Corresponding author: \href{mailto:chengjun@huas.edu.cn}{chengjun@huas.edu.cn}}

\author{En-Kun Li}
%\email
\thanks{Corresponding author: \href{mailto:lienk@mail.sysu.edu.cn}{lienk@mail.sysu.edu.cn}}

\author{Jianwei Mei}
%\email
\thanks{Corresponding author: \href{mailto:meijw@mail.sysu.edu.cn}{meijw@mail.sysu.edu.cn}}

\affiliation\TRC
\date{\today}%
%
%
%%%%%%%%%%%%%%%%%%%%%%%%%%%%%%%%%%%%%%%%%%%%%%%%%%%%%%%%%%%%%%%%%%%%%%%%%%
\begin{abstract}
%%Here goes the abstract \nocite{*}%% Remove this line from your manuscript.
    The stochastic gravitational wave background (SGWB) is one of the main detection targets for future millihertz space-based gravitational-wave observatories such as the \ac{LISA}, TianQin, and Taiji.
For a single LISA-like detector, a null-channel method was developed to
identify the SGWB by integrating data from the A and E channels with
a noise-only T channel.
However, the noise monitoring channel will not be available if one of the laser interferometer arms fails.
By combining these detectors, it will be possible to build detector networks to search for SGWB via cross-correlation analysis.
In this work, we developed a Bayesian data analysis method based on
\ac{TDI} Michelson-type channel.
We then investigate the detectability of the TianQin-LISA detector network for various isotropic SGWB.
Assuming a three-month observation, the TianQin-LISA detector network could be able to confidently detect SGWB with energy density as low as
$\Omega_{\rm PL} = 6.0 \times 10^{-13}$, $\Omega_{\rm Flat} = 2.0 \times 10^{-12}$
and $\Omega_{\rm SP} = 1.2 \times 10^{-12}$ for power-law, flat and single-peak models, respectively.
\end{abstract}
%%%%%%%%%%%%%%%%%%%%%%%%%%%%%%%%%%%%%%%%%%%%%%%%%%%%%%%%%%%%%%%%%%%%%%%%%%

\maketitle

\acrodef{GW}{gravitational wave}
\acrodef{GCB}{Galactic ultra compact binaries}
\acrodef{SGWB}{Stochastic gravitational wave background}
\acrodef{SNR}{signal-to-noise ratio}
\acrodef{DWD}{double white dwarf}
\acrodef{MBHB}{massive black hole binary}
\acrodefplural{MBHB}[MBHBs]{massive black hole binaries}
\acrodef{SBBH}{stellar-mass black hole binary}
\acrodefplural{SBBH}[SBBHs]{stellar-mass black hole binaries}
\acrodef{EMRI}{extreme mass ratio inspiral}
\acrodef{PTA}{pulsar timing array}
\acrodef{CE}{cosmic explorer}
\acrodef{ET}{Einstein telescope}
\acrodef{LISA}{laser interferometer space antenna}
\acrodef{O1}{first observing run}
\acrodef{O3}{third observing run}
\acrodef{LVK}{the LIGO Scientific Collaboration, the Virgo Collaboration and the KAGRA Collaboration}
\acrodef{PCA}{principal component analysis}
\acrodef{CBC}{compact binary coalescences}
\acrodef{TDI}{time delay interferometry}
\acrodef{PSD}{power spectral density}
\acrodef{SNR}{signal-to-noise ratio}
\acrodef{PDF}{probability distribution function}
\acrodef{MCMC}{Markov Chain Monte Carlo}
\acrodef{NS}{nested sampling}
\acrodef{PTAs}{pulsar timing arrays}

% Main text
\section{Introduction}
\label{sec:intro}

The direct observation of the first gravitational-wave event, GW150914, resulting from the merger of a binary black hole system \citep{LIGOScientific:2016aoc}, has opened up a whole new era of observing the Universe. %in the observation of the Universe.
Since then, an increasing number of \acp{GW} from
\ac{CBC}, nearly a hundred to date, have been detected by LIGO and Virgo \citep{LIGOScientific:2016sjg,LIGOScientific:2017vwq,LIGOScientific:2018mvr,LIGOScientific:2020aai,LIGOScientific:2020ibl,LIGOScientific:2021qlt}.
In addition, there are a large number of GWs that cannot be resolved individually, leading to the formation of \ac{SGWB} \citep{Romano:2016dpx,Christensen:2018iqi}.
\ac{SGWB} carries important information about the early Universe and astrophysical source populations, %and its detection
and detecting such a background signal
would have a significant impact on fundamental physics and astrophysics \citep{Mazumder:2014fja,Callister:2016ewt,Maselli:2016ekw}.

\ac{SGWB} is extremely weak and is often overwhelmed by Galactic foreground and detector noise.
%Indeed, identifying a \ac{SGWB} is challenging due to its statistical properties are indistinguishable from instrument noise.
Indeed, it is challenging to identify \ac{SGWB} due to its statistical properties are indistinguishable from instrument noise.
However, the cross-correlation method \citep{Hellings:1983fr,Christensen:1992wi,Flanagan:1993ix,Allen:1997ad}
provides a powerful tool for the detection of \ac{SGWB} when multiple detectors are available.
This method is based on the principle that \ac{SGWB} recorded by multiple detectors is correlated, while noise from different detectors is statistically independent.

%Ground-based GW detectors currently in operation, including Advanced LIGO \cite{LIGOScientific:2014qfs}, Advanced Virgo \cite{VIRGO:2014yos} and KAGRA \cite{Somiya:2011np} in the hertz sensitive band, as well as \ac{PTAs} \cite{Detweiler:1979wn} in the nano-hertz sensitive band, all utilize cross-correlation method to search for a \ac{SGWB}.

Ground-based GW detectors currently in operation, including Advanced LIGO \citep{LIGOScientific:2014qfs}, Advanced Virgo \citep{VIRGO:2014yos}, and KAGRA \citep{Somiya:2011np}, which are sensitive in the hertz band,
as well as \ac{PTAs} \citep{Detweiler:1979wn} sensitive in the nano-hertz band, all utilize the cross-correlation method to search for a \ac{SGWB}.
For example, the \ac{LVK} has established upper bounds on the fractional energy density spectrum $\Omega_{\rm gw} \le 5.8 \times 10^{-9}$ of a frequency-independent (flat) \ac{SGWB} \citep{KAGRA:2021kbb}.
What's more, \ac{PTAs} have now measured the Hellings owns correlation with high significance, providing convincing evidence for the existence of \ac{SGWB}
\citep{NANOGrav:2023gor,EPTA:2023fyk,Xu:2023wog,Reardon:2023gzh}.

In the mHz band, space-based missions like \ac{LISA} \citep{LISA:2017pwj}, TianQin \citep{Luo:2015ght}, and Taiji \citep{Hu:2017mde}, are underway.
Such detectors promise richer GW sources,
including the inspiral of Galactic ultra-compact binaries (GCBs) \citep{Huang:2020rjf},
the merger of massive black hole binaries (MBHBs) \citep{Wang:2019ryf},
the extreme mass ratio inspirals (EMRIs) \citep{Zi:2024itp, Zi:2024mbd, Zi:2023geb,Zi:2021pdp,Fan:2020zhy},
the inspiral of stellar-mass black hole binaries (SBHBs) \citep{Liu:2020eko,Liu:2021yoy},
and potentially energetic processes in the early
Universe \citep{Cheng:2022vct,Liang:2021bde,vanDie:2024htf}.

For an individual \ac{LISA}-type, although multiple channels can be constructed through
the \ac{TDI} \citep{Tinto:1999yr,Prince:2002hp,Tinto:2020fcc},
the cross-correlation method is not available due to correlated noise
between the channels \citep{Cheng:2022vct,Adams:2010vc,Adams:2013qma,Wang:2022sti}.
The currently proposed null-channel method for \ac{SGWB} detection relies on the combined use of three TDI channels,
including one that is insensitive to GWs, such as the T-channel \citep{Adams:2010vc}.
For this approach to be effective, it is essential to understand the relationship between the instrumental noise \ac{PSD} of the T-channel and that of the GW channels.
Furthermore, the T-channel is constructed from a linear combination of the output of the three laser interferometric arms.
If any one of these arms fails, the null-channel method will not be available.
However, by combining multiple (two or more) space-based GW detectors,
it will be feasible to build detector networks capable of the detection of \ac{SGWB}.% in the future.

Numerous interesting work has been carried out on \ac{SGWB} detection
with the space-based \ac{GW} detector networks \citep{Seto:2020zxw,Omiya:2020fvw,Seto:2020mfd,Orlando:2020oko,Wang:2021njt,Wang:2022pav,Wang:2023ltz,Hu:2023nfv,Hu:2024toa}.
Among them, Ref.~\citep{Seto:2020zxw} presents the first quantitative analysis for measuring parity asymmetry in the \ac{SGWB} by utilizing data streams from the LISA-Taiji detector network.
Ref.~\citep{Omiya:2020fvw} further investigated the detectability of anomalous polarization modes of \ac{SGWB}.
More general detector networks have been proposed to search for isotropic background signals \citep{Seto:2020mfd}.
Additionally, Bayesian data analysis methods based on the LISA-TianQin network have also been developed \citep{Hu:2023nfv,Hu:2024toa}.
Conversely, scant data analysis efforts have been devoted to directly evaluating the parameter estimates and detection limits of the \ac{SGWB} as of yet.

In this work, we formulate a Bayesian data-analysis framework for \ac{SGWB} searches that  exploits the two independent Michelson-type channels available to networks of space-based laser interferometers.
We implement the formalism for the prospective TianQin-ISA network and evaluate its detectability for various \ac{SGWB} energy-density spectra.
Unlike ground-based GW detector networks, the relative positions of detectors within space-based detector networks will vary throughout their operational lifetime.
As a result, the overlap reduction functions (ORFs) \citep{Finn:2008vh},
being a time-variant quantity, pose new problems and challenges in the detection of \ac{SGWB}.
We therefore divide the total observation data into several segments so that the ORFs change very little within each short time-series segment, allowing us to treat them as nearly constant.
Ultimately, these data segments are generally equivalent to data from multiple detector pairs observed at different times, similar to the ground-based detector networks.

This paper is organized as follows.
In Sec. \ref{sec:SGWB}, we introduce the basic properties and mathematical description of various \ac{SGWB} models.
In Sec. \ref{sec:DETECTOR NETWORK}, we describe in detail the orbital characteristics and instrument noise model of the TianQin-LISA joint network.
The Sec. \ref{sec:SEARCH METHOD} briefly reviews the cross-correlation analysis, obtains the time-varying ORFs at various times,
and presents a Bayesian analysis technique.
In Sec. \ref{sec:Results}, we show the main results of the TianQin-LISA detector network for detecting various background signals.
Finally, conclusions and discussions of possible extensions of this work are drawn in Sec. \ref{sec:SUMMARY AND DISCUSSION}.

%%%%%%%%%%%%%%%%%%%%%%%%%%%%%%%%%%%%%%%%%%%%%%%%%%%%
\section{Description of the SGWB}
\label{sec:SGWB}

\ac{SGWB} is a random signal formed by the superposition of many weak GWs, and its metric perturbations can be written as the superposition of plane waves propagating in the transverse-traceless gauge:
\beq
\label{expa}
h_{ij}(t,\vec{x}) =
\int_{-\infty}^{\infty} df
\int_{S^{2}} d{\hat{k}} \, h_{ij}(f,\hat{k})\, e^{ {\rm i}2\pi f(t-\hat{k}\cdot \vec{x}/c)}.
\eeq
Here, $h_{ij}(f,\hat{k})= \sum_{P} \, h_{P}(f,\hat{k}) \, e_{ij}^P(\hat{k})$,
$\hat{k}$ stands for the direction of the GW propagation,
$h_{P}(f,\hat{k})$ denotes Fourier components,
$P \in \{+,\times\}$ represents polarization states,
and $e_{ij}^P$ are the polarization tensors.

For a Gaussian, stationary, unpolarized, and isotropic \ac{SGWB}, the statistical properties of its Fourier components are completely characterized by their second-order moments, which satisfy the following relations \citep{Romano:2016dpx,Allen:1997ad}:
\beq
\label{eq:sh}
\left\langle h_{P}(f,\hat k) h_{P'}^*(f',\hat k') \right\rangle
= \frac{S_h(f) \delta_{PP'}\delta(f-f')  \delta^2(\hat k,\hat k')}{16\pi}  ,
\eeq
where angle brackets $\langle ... \rangle$ denotes the ensemble average,
$ \delta^2(\hat k, \hat k') := \delta(\phi - \phi') \, \delta(\cos\theta - \cos\theta')$ is the covariant Dirac delta function on the two-sphere
, and $S_h(f)$ is the one-sided \ac{PSD}.
On the other hand, $S_h(f)$ is related to the fractional energy density spectrum $\Omega_{\text{gw}}(f)$ of the \ac{SGWB} as
\beq
S_h(f)=\frac{3H_0^2}{2\pi^3} \frac{\Omega_{\text{gw}}(f)}{f^{3}},
\text{ and }
\Omega_{\text{gw}} (f) = \frac{1}{\rho_c} \frac{d\rho_{\text{gw}}}{d\ln f}.
\eeq
Here, $\rho_c = 3H_0^2/(8\pi G)$ is the critical energy density of the Universe,
$H_0$ is the Hubble constant,
$G$ represents the gravitational constant, and $\rho_{\rm gw}$ denotes the GW energy density.

Several theoretical models predict that the \ac{SGWB} spectrum exhibits a power-law behavior \citep{Regimbau:2011rp,Farmer:2003pa,Moore:2014lga}, described by
\beq\label{eq:PLspectrum}
\Omega_{\text{gw}}(f) = \Omega_{\text {n}} \, \left( \frac{f}{ f_{\text{ref}} } \right)^{n} ,
\eeq
where $\Omega_{\text{n}}$ represents the amplitude level at the reference frequency $f_{\text{ref}}$, and $n$ is spectral index.
For instance, in astrophysical scenarios like \ac{CBC}, the \ac{SGWB} spectrum commonly assumed that
\beq
\Omega_{\text{CBC}}(f) \propto f^{2/3},
\eeq
corresponding to a spectral index of $n = 2/3$.
In contrast, the \ac{SGWB} predicted by slow-roll inflation scenarios \citep{Grishchuk:1974ny,Grishchuk:1993ds,Starobinsky:1979ty,Maggiore:1999vm}
is characterized by a flat spectrum in the TianQin and LISA detection frequency regime,
\beq
\label{eq:Flatspectrum}
\Omega_{\text{Inflation}}(f) = \Omega_{\text{Flat}},
\eeq
with $n = 0$.
Another physical mechanism in the very early Universe could produce
SGWB with a Gaussian-bump (or single-peak) phenomenological spectrum \citep{Thorne:2017jft,Namba:2015gja},
\beq\label{eq:SPspectrum}
\Omega_{\text{Bump}} = \Omega_{\text{SP}} \exp\left\{-\frac{[\log_{10}(f/f_{\text{ref}})]^2}{\Delta^2}\right\}.
\eeq

Throughout the work, we adopt the reference frequency to $f_{\text{ref}} = 1$ mHz for the power-law spectrum and $f_{\text{ref}} = 3$ mHz for the single-peak spectrum.
The fiducial value of the Hubble constant, $H_0 = 67 \, \text{km s}^{-1} \text{Mpc}^{-1}$, is consistent with the Planck 2015 result \citep{Planck:2015fie}.

%%%%%%%%%%%%%%%%%%%%%%%%%%%%%%%%%%%%%%%%%%%%
\section{Detector Network}
\label{sec:DETECTOR NETWORK}

\begin{figure}[htbp]
\centering
\includegraphics[width=\linewidth]{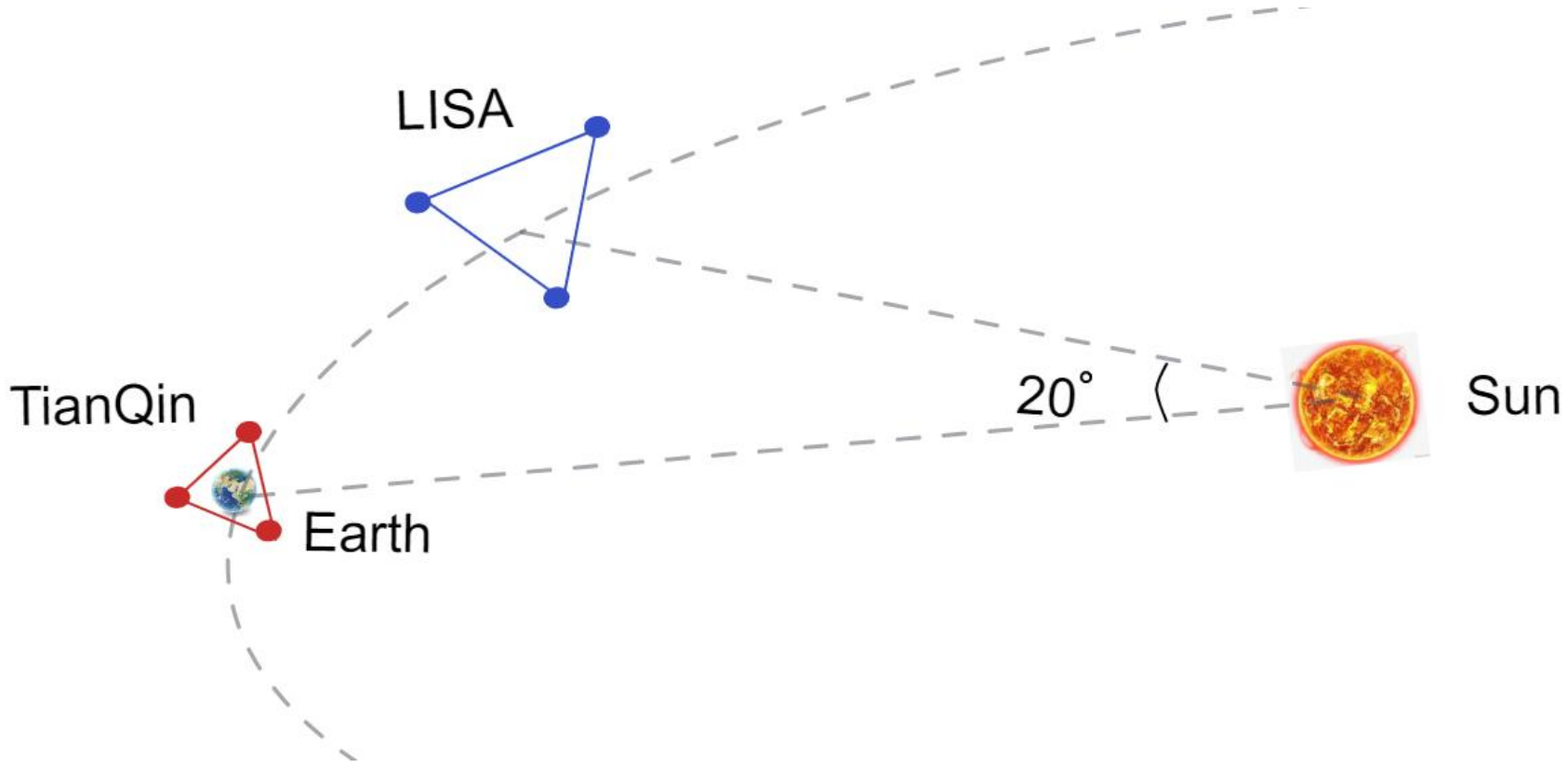}
\caption{Illustration of the TianQin-LISA detector network.
TianQin is expected to operate in a geocentric orbit,
whereas LISA is designed to orbit the Sun, trailing $20^\circ$ behind the Earth.}
\label{fig:Network}
\end{figure}

Both TianQin and \ac{LISA} are space-based \ac{GW} observatories designed to detect \ac{GW} sources in the mHz frequency band,
which will reveal richer astrophysical phenomena and provide complementary and validation for ground-based \ac{GW} detectors.
Each observatory will consist of three drag-free satellites in an equilateral triangular configuration,
and has a detector sensitivity to $[\,10^{-4},\,1\,]$ Hz, as shown in Fig. \ref{fig:Network}.

\subsection{Orbital characteristics}

%LISA, in a heliocentric orbit $20^\circ$ behind Earth, has its satellite orbits in the ecliptic coordinate system given by:
LISA is expected to be positioned in a heliocentric orbit \citep{LISA:2017pwj},
trailing Earth by $20^\circ$,
with the arm-length $L_{\text{LISA}} = 2.5 \times 10^{9}$ m,
and its satellite's orbits are defined within the ecliptic coordinate system as follows:
\begin{align}
\nn
X_n &= R \cos \alpha + \frac{e R}{2}
\left[ \cos(2\alpha - \kappa_n) - 3\cos \kappa_n \right] + \mathcal{O}(e^2),
\\
\nn
Y_n &= R \sin \alpha + \frac{e R}{2}
\, \left[\, \sin(2\alpha - \kappa_n) - 3\sin \kappa_n \right] + \mathcal{O}(e^2), \\
Z_n &= -\sqrt{3}\,e R \cos(\alpha - \kappa_n) + \mathcal{O}(e^2).
\end{align}
Here, $R = 1 \text{AU}$, $\alpha = 2\pi f_{m} t - \beta$, $ f_{m} = 1/\text{yr}$,
$e = 0.0048$, $\kappa_n = \frac{2\pi (n-1)}{3}$, and $n = 1, 2, 3$ corresponds to each satellite in the LISA constellation.
%to each of the three satellites in the LISA constellation.
Higher-order eccentricity terms, $\mathcal{O}(e^2)$, are neglected in this analysis.

%%%%%%%%%%%%%%%%%%%%%%%%%%%%%%%%%%%%%%%%%%%%%%%%%%%%%%%%%%%%%%%%%%%%%%%%%%
%%%%%%%%%%%%%%%%%%%%%%%%%%%%%%%%%%%%%%%%%%%%%%%%%%%%%%%%%%%%%%%%%%%%%%%%%%

%Unlike the LISA mission, TianQin will adopt
In contrast to LISA, TianQin operates in a geocentric orbital with an approximately constant detector orientation,
and features a shorter laser interferometry arm,
$L_{\text{TQ}} = \sqrt{3} \times 10^{8}$ m \citep{Luo:2015ght}.
%for laser interferometry measurements.
Furthermore, in order to avoid the effects of Earth and Moon eclipses and also fulfill the constellation stability requirements,
TianQin operates in a 3 + 3 operation mode, that is, three months on, three months off \citep{Ye:2019txh,Zhang:2020paq}.

The ecliptic coordinates of the TianQin satellite system consist of two components: the Earth's ecliptic coordinates $(X_n^{\text{TQ}}, Y_n^{\text{TQ}}, Z_n^{\text{TQ}})$,
and the geocentric ecliptic coordinates $( x_n^{\text{TQ}}, y_n^{\text{TQ}}, y_n^{\text{TQ}})$. The former are described by:
\begin{align}
\label{eq:cotq1}
\nn
X_n^{\text{TQ}} &= R \cos(\alpha^{\text{TQ}}) + \frac{1}{2} e^{\text{TQ}} R \left[ \cos(2\alpha^{\text{TQ}}) - 3 \right], \\
\nn
Y_n^{\text{TQ}} &= R \sin(\alpha^{\text{TQ}}) + \frac{1}{2} e^{\text{TQ}} R \sin(2\alpha^{\text{TQ}}), \\
Z_n^{\text{TQ}} &= 0,
\end{align}
with $e^{\text{TQ}} = 0.0167$, $\alpha^{\text{TQ}} = 2\pi f_{\text{m}} t - \beta +20^\circ$, and $\beta = 102.9^\circ$ is the longitude of perihelion.
The geocentric ecliptic coordinates for TianQin, relative to the Earth's center, are written as
\begin{align}
\nn
x_n^{\text{TQ}} &= R_1 \biggl(\cos\phi_{\text{s}} \sin\theta_{\text{s}} \sin \alpha_n + \cos\alpha_n \sin\phi_{\text{s}}\biggr), \\
\nn
y_n^{\text{TQ}} &= R_1 \biggl(\sin\phi_{\text{s}} \sin\theta_{\text{s}} \sin\alpha_n - \cos\alpha_n \cos\phi_{\text{s}}\biggr), \\
z_n^{\text{TQ}} &= -R_1 \sin\alpha_{n} \cos\theta_{\text{s}},
\end{align}
where $R_1 = 1 \times 10^8 \, \text{m}$,
$\alpha_n = 2\pi f_{\text{sc}} t + \kappa_n$, with $f_{\text{sc}} = 1/(3.64 \, \text{days})$,
$\theta_{\text{s}} = -4.7^\circ$, and $\phi_{\text{s}} = 120.5^\circ$.

%%%%%%%%%%%%%%%%%%%%%%%%%%%%%%%%%%%%%%%%%%%%%%%%%%%%%%%%%%%%%%%%%%%%%%%%%%
\begin{figure}[htbp]
\centering
\includegraphics[width=0.7\linewidth]{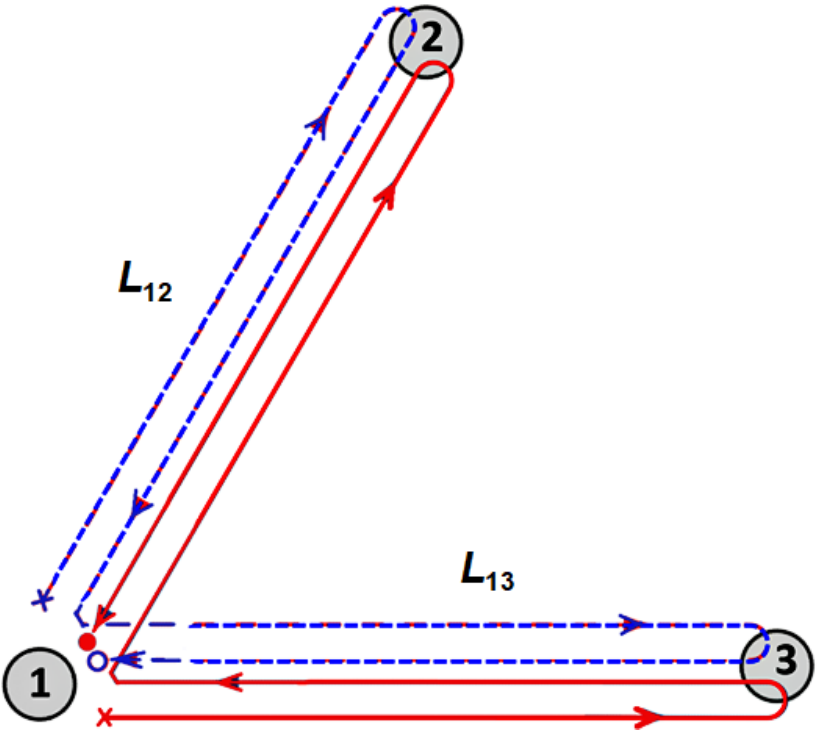}
\caption{Schematic diagram of the TDI-1.5 generation Michelson-type X channel.}
\label{fig:MichelsonX}
\end{figure}
%%%%%%%%%%%%%%%%%%%%%%%%%%%%%%%%%%%%%%%%%%%%%%%%%%%%%%%%%%%%%%%%%%%%%%%%%%

\subsection{Noise model}
Laser links will be established among the three satellites comprising the detector,
facilitating the detection of GWs through laser interferometry.
This technique enables the precise measurement of minuscule changes in the distance between free-falling test masses due to the passage of GWs.

Consider satellite $\text{SC}_i$ transmitting a laser pulse to satellite $\text{SC}_j$. The time series of the laser phase $\Phi_{ij}(t)$ recorded by satellite $\text{SC}_j$ is composed of instrument noise and GWs $\psi_{ij}(t)$, which can be expressed as:
\beqa
\Phi_{ij}(t) &=&
C_i\left(t - \frac{L_{ij}}{c}\right) - C_j(t) + \psi_{ij}(t) + n_{ij}^p(t)
\nonumber \\
&-& \hat{r}_{ij} \cdot \left[ \vec{n}_{ij}^a(t) - \vec{n}_{ji}^a\left(t - \frac{L_{ij}}{c}\right) \right].
\eeqa
Here, $C(t)$ represents laser frequency noise, $n_{ij}^p(t)$ denotes displacement noise,
$\vec{n}^a(t)$ signifies acceleration noise, $\hat{r}_{ij}$ is the unit vector pointing from satellite $\text{SC}_i$ to satellite $\text{SC}_{j}$,
$L_{ij}$ is the arm-length between satellites $\text{SC}_i$ and $\text{SC}_j$, and $c$ is the speed of light.

Laser frequency noise usually dominates the observed data, several orders of magnitude higher than other noise.
To suppress laser frequency noise effectively, \ac{TDI} is used in the process of data
post-processing \citep{Tinto:1999yr,Prince:2002hp,Tinto:2020fcc}.
In this work, we focus our analysis on the TDI-1.5 generation Michelson-type channel.

With the output phases from six different links, it is possible to construct three virtual equal-arm channels, which are referred to as \{X, Y, Z\} channels.
For instance, the Michelson-X channel (shown in Fig. \ref{fig:MichelsonX}) can be built as follows:
\beqa
%\label{XXpsd}
\text{X}(t) &= & \left[\Phi_{12}(t - 3L/c) + \Phi_{21}(t - 2L/c)\right]
\nonumber \\  &&
+ \left[\Phi_{13}(t - L/c) + \Phi_{31}(t)\right]
\nonumber \\ &&
- \left[\Phi_{13}(t - 3L/c) + \Phi_{31}(t - 2L/c)\right]
\nonumber \\ &&
- \left[\Phi_{12}(t - L/c) + \Phi_{21}(t)\right].
\eeqa
The Fourier transform of the variable $X(t)$ is
\beqa
\label{TDI-1.5X}
\text{X}^a(f) &= & 4\text{i} \sin u e^{-\text{i}2u} \left[\left(n_{31}^a + n_{21}^a\right) \cos u - \left(n_{12}^a + n_{13}^a\right)\right],
\nonumber \\
\text{X}^p(f) &= & 2\text{i} \sin u e^{-\text{i}2u} \left[\left(n_{31}^p - n_{21}^p\right) e^{\text{i}u} + \left(n_{13}^p - n_{12}^p\right)\right],
\nonumber \\
\eeqa
where $u = f/f_*$, and $f_* = \frac{c}{2\pi L}$ is the characteristic frequency.
The other Michelson-type channels, Y and Z, can be derived by cyclically permuting the indices $1 \rightarrow 2 \rightarrow 3 \rightarrow 1$ in Eqs. (\ref{TDI-1.5X}).

Assuming that all arm lengths are equal, $L_{ij}=L$,
and that the displacement noise and acceleration noise of each satellite are completely symmetric,
the corresponding channel noise \ac{PSD} can be analytically expressed as
\beq
\label{XXpsd}
\left\langle \text{XX}^*\right\rangle = 16 \sin^2 u
\left[ S_{p}+ 2 S_{a}\left( 1 + \cos^2 u \right) \right].
\eeq
Here,
$S_{p} \equiv S_{ij}^{p}= \left\langle n^p_{ij}\,n^{p*}_{ij}\right \rangle$,
and
$S_{a} \equiv S_{ij}^{a}= \left \langle n^a_{ij}\,n^{a*}_{ij}\right\rangle$
are the nominal spectral density of displacement noise and acceleration noise respectively,
\beqa
\label{tqspsa}
S^{\rm tq}_{p} &=&
N^{\rm tq}_{p}\left[\frac{\mathrm{m}^{2}}{\mathrm{Hz}} \right]
\left(\frac{1}{2L}\right)^{2},
\nonumber \\
S^{\rm lisa}_{p} &=&
N^{\rm lisa}_{p} \left[ \frac{\mathrm{m}^{2}}{\mathrm{Hz}} \right]
\left(\frac{1}{2L}\right)^{2}\left[ 1+ \left(\frac{2 \mathrm{mHz}}{f} \right)^4 \right],
\nonumber \\
S^{\rm tq}_{a} &=&
N^{\rm tq}_{a} \left[\frac{\mathrm{m}^{2}}{\mathrm{s}^{4}\,\mathrm{Hz}}\right]
\left(\frac{1}{2L}\right)^{2} \left(\frac{1}{2\pi f}\right)^4
\left(1+ \frac{0.1 \mathrm{mHz}}{f}  \right),
\nonumber \\
S^{\rm lisa}_{a} &=&
N^{\rm lisa}_{a} \left[ \frac{\mathrm{m}^{2}}{\mathrm{s}^{4}\,\mathrm{Hz}}\right]
\left(\frac{1}{2L}\right)^{2} \left(\frac{1}{2\pi f}\right)^4
\left(1+ \frac{0.4 \mathrm{mHz}}{f} \right)
\nonumber \\
&&
\times \left[1+ \left( \frac{f}{8 \mathrm{mHz}} \right)^4  \right].
\eeqa

\begin{figure}[htbp]
\centering
\includegraphics[width=\linewidth,trim={0.1cm 0 1.2cm 1cm},clip]{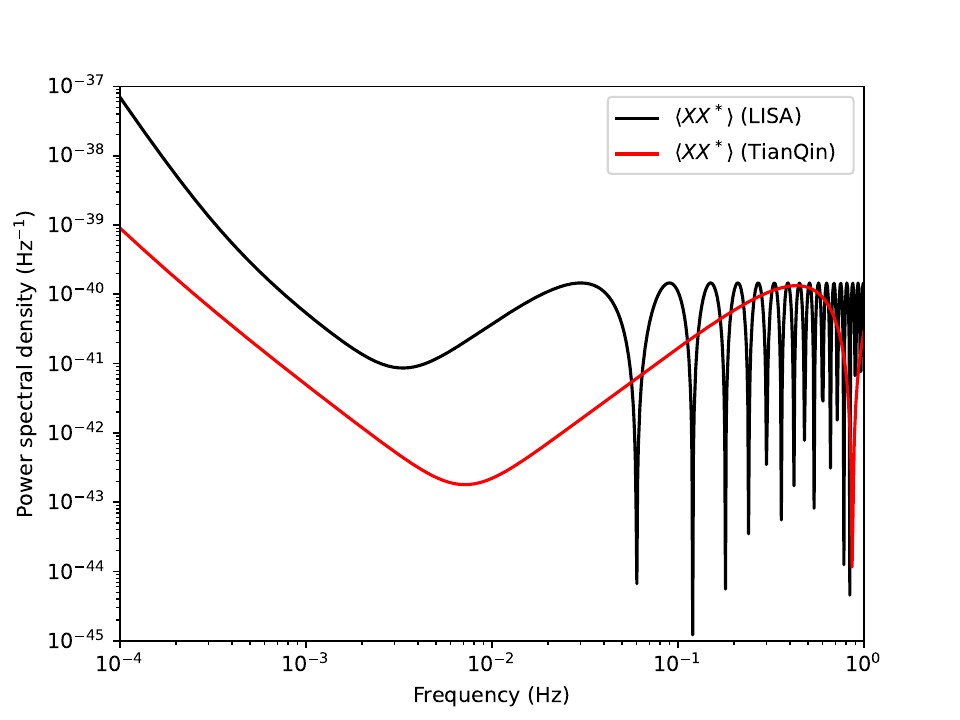}
\caption{The TDI-1.5 generation Michelson-X channel noise \ac{PSD} from different space-based GW detectors.}
\label{fig:XPSD}
\end{figure}

The key parameters and instrument noise \ac{PSD} of the Michelson-X channel
from different mission configurations are shown in Table~\ref{tab:Parameter}
and Fig. \ref{fig:XPSD}, respectively.

%%%%%%%%%%%%%%%%%%%%%%%%%%%%%%%%%%%%%%%%%%%%%%%%%%%%%%%%%%%%%%%%%%%%%%%%%%
\begin{table}[t]
    \centering
    \caption{Key parameters for the TianQin and LISA configurations.}
    \label{tab:Parameter}
    \begin{tabular}{c|c|c}
        \toprule
        Parameter & \text{TianQin} &  LISA \\ %& With image \\ \mathrm{~m}^{2} \mathrm{~s}^{-4}
        \midrule
         $L \left[\mathrm{m}\right]$  & $\sqrt{3} \times 10^{8}$                    & $2.5 \times 10^{9}$ \\
         $N_{p} \left[\mathrm{m}^{2} / \mathrm{Hz}\right]$  & $1.0 \times 10^{-24} $    & $ 2.25 \times 10^{-22}$ \\
         $N_{a} \left[\mathrm{m}^{2} / \mathrm{s}^{4} /\mathrm{Hz}\right]$  & $1.0 \times 10^{-30} $ & $9.0 \times 10^{-30}$ \\
         %\text {Orientation }  & $\lambda=120.4^{\circ}$, $\beta=-4.7^{\circ}$ & $\lambda=120.4^{\circ}$, $\beta=-4.7^{\circ}$ \\
         \text {Operation } & $ 2 \times 3  $ \text{months each year} & \text{Year-round } \\
        \bottomrule
    \end{tabular} %\\ \parbox[t]{2.0 \linewidth}{\flushleft\footnotesize ins. }
\end{table}
%%%%%%%%%%%%%%%%%%%%%%%%%%%%%%%%%%%%%%%%%%%%%%%%%%%%%%%%%%%%%%%%%%%%%%%%%%

%%%%%%%%%%%%%%%%%%%%%%%%%%%%%%%%%%%%%%%%%%%%%%%%%%%%%%%%%%%%%%%%%%%%%%%%%%
%%%%%%%%%%%%%%%%%%%%%%%%%%%%%%%%%%%%%%%%%%%%%%%%%%%%%%%%%%%%%%%%%%%%%%%%%%

\section{Search method}
\label{sec:SEARCH METHOD}

The background signal is very weak, and often acts as another source of noise in a single detector, especially in a single space-based \ac{GW} detector \citep{LISA:2017pwj,Luo:2015ght,Hu:2017mde}.
For the special case of a triangular-shaped \ac{GW} detector,
the cross-correlation method now widely used in ground-based detector networks will not be available \citep{Adams:2010vc,Adams:2013qma,Wang:2022sti,Cheng:2022vct}.
However, the recent rapid development of mHz space-based \ac{GW} detectors
has made it possible to construct space-based \ac{GW} detector networks for identifying \ac{SGWB} from complex noise via cross-correlation analysis \citep{Seto:2020zxw,Omiya:2020fvw,Seto:2020mfd,Orlando:2020oko,Wang:2021njt,Wang:2022pav,Wang:2023ltz,Hu:2023nfv,Hu:2024toa}.

This section reviews the cross-correlation method and derives the fundamental expressions necessary for the TianQin-LISA detector network.
%%%%%%%%%%%%%%%%%%%%%%%%%%%%%%%%%%%%%%%%%%%%%%%%%%%%%%%%%%%%%%%%%%%%%%%%%%

\subsection{Correlation analysis}
%\label{sec:Detector}

The output of each channel in the TianQin-LISA detector network is typically composed of signal and instrument noise,
%In general, the output of each channel in an interferometer consists of the instrumental noise and the signal,
\begin{align}
%\beq
d_{I}(f) =& \, h_{I}(f)+ n_{I}(f),   %I = \{A, E, T\}  ,
\nonumber \\
d_{J}(f) =& \, h_{J}(f)+ n_{J}(f),   %I = \{A, E, T\}  ,
%\eeq
\end{align}
where $n(f)$ represents the Fourier transform of the detector noise, and
\beq
\label{eq:responsesig}
h(f)=\sum_{P} \int_{S^{2}} d{\hat{k}} \, h_{P}(f,\hat{k})F^{P}(f,\hat{k})
e^{-{\rm i}2\pi f\hat k\cdot \vec x /c},\,
\eeq
is the interferometric response to the SGWB, with antenna pattern function $F^{P}(f,\hat{k})$.

Assuming that there is no correlation between the instrument noise in each detector,
\beq
%\label{eq:}
\langle n_{I}(f) \, n_{J}^*(f\rq{})\rangle =
\begin{cases}
~~~~~~~~~~~~~~0~~~~~~~~, & I \neq J \\
~\frac{1}{2} \, \delta(f-f\rq{}) \, P_{n}(f) ,& I =J
\end{cases}
\eeq
where $ P_{n}(f)$ is the PSD of the instrument noise,
the background signal can then be extracted by cross-correlating the outputs of the detector network,
\beq
\langle d_{I}(f)\, d_{J}^*(f\rq{})\rangle
=
\frac{1}{2}\,\delta(f-f\rq{})\,{\Gamma}_{IJ}(f)\,S_h(f).
\eeq
Here, $\Gamma_{IJ}(f)$ is the ORFs \citep{Finn:2008vh},
which quantifies the correlated responses of the detector network to the \ac{SGWB}.

\subsection{Overlap reduction function}

The general expression of $\Gamma_{IJ}(f)$ can be derived from Eqs.~(\ref{eq:sh}) and (\ref{eq:responsesig}),
\beq
\label{eq:ORF}
\Gamma_{IJ}(f) = \frac{1}{8\pi} \sum_{P}\int_{}F_I^P(f,\hat k) F_{J}^{P*}(f,\hat k) e^{ -{\rm i}2\pi f\hat{k} \cdot \Delta \vec{x}/c} d{\hat k},
\eeq
where $\Delta \vec{x}  = \vec{x}_I(t) - \vec{x}_J(t)$ represents the relative separation vector of detectors $I$ and $J$,
with $\vec{x}_{I}(t)$ and $\vec{x}_J(t)$ being their positions at time $t$.
In ground-based \ac{GW} detector networks, $\Gamma_{IJ}(f)$ is time-invariant due to the fixed relative positions $\Delta \vec{x}$,
and thus has analytical expressions in the low-frequency approximations \citep{Flanagan:1993ix,Allen:1997ad}.
For space-based \ac{GW} detector networks, however, the ORFs are expected to exhibit dynamic behavior,
and the small antenna approximation may not consistently hold within their sensitive frequency bands \citep{Hu:2023nfv,Hu:2024toa,Liu:2022umx}.

\begin{figure*}[htbp]%[!ht]!htbp
    \includegraphics[width=0.48\linewidth,clip]{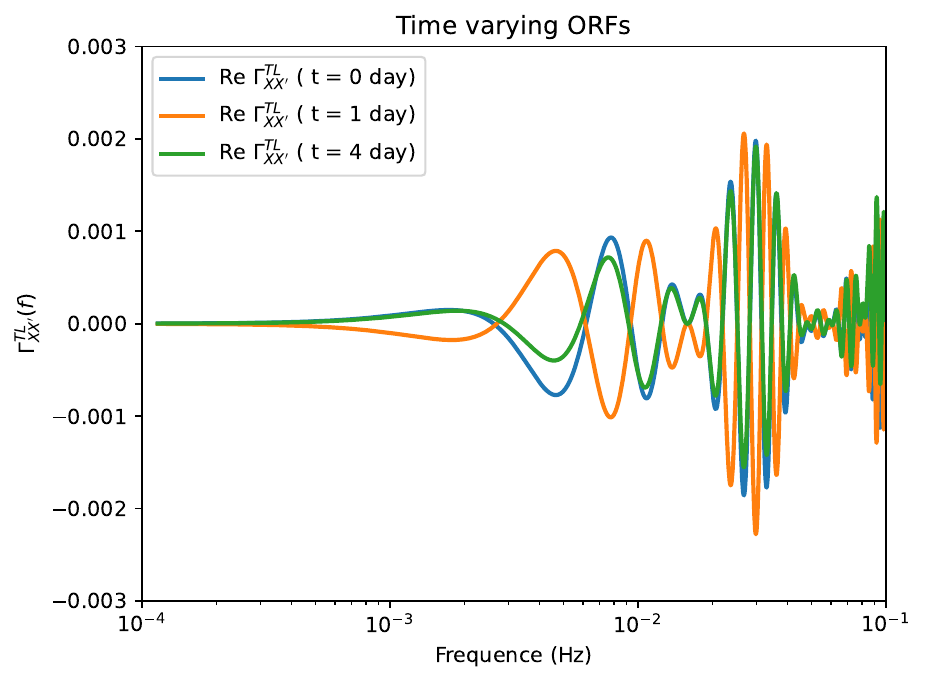}
    \includegraphics[width=0.48\linewidth,clip]{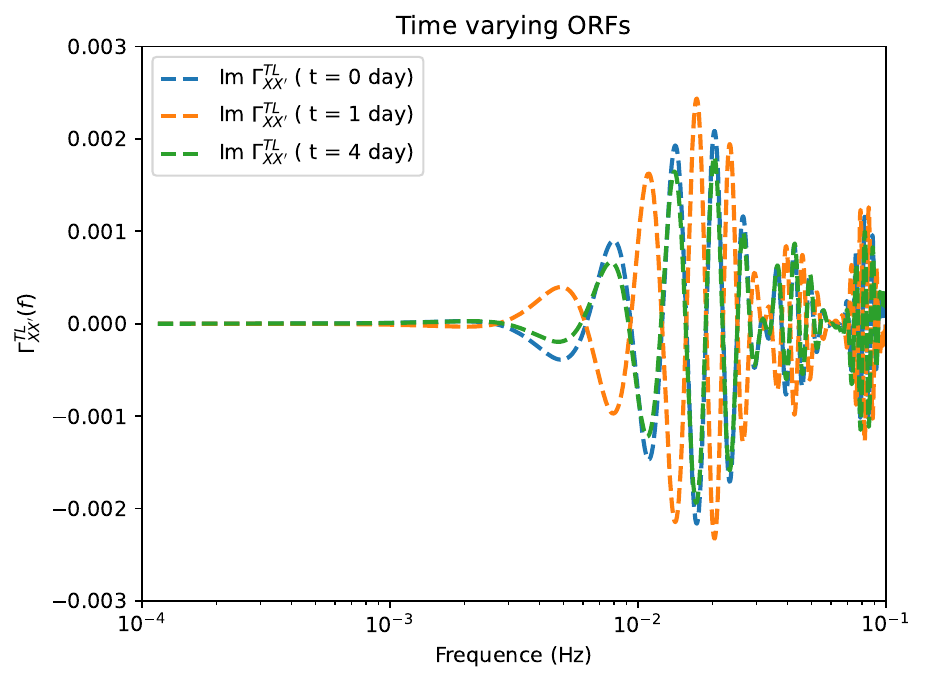}
   \caption{The overlap reduction function as a function of time in X channel,
    which shows the detector network's correlated responses to the \ac{SGWB} at $t = 0$, $t = 1$ day, and $t = 4$ day.
    The upper panel illustrates the real parts, while the lower panel represents the imaginary parts.}
\label{fig:ORF}
\end{figure*}

We utilize numerical methods to calculate the Michelson-type X channel ORFs, $\Gamma_{\text{XX}'}(f)$.
The time-dependent nature of $\Gamma_{\text{XX}'}(f)$ is illustrated in Fig. \ref{fig:ORF}, with the upper and lower panels showing the evolution of the real and imaginary parts, respectively.
The ORFs vary greatly over time in terms of a day.
These changes in the ORFs could potentially impact the joint network's detection capability, including parameter estimation and detection limits.
To address this, we divide the total observation data into segments of 0.2 hours duration.
Within each segment, the variation in the ORFs is small enough to be approximated as constant, allowing for a piecewise cross-correlation analysis.
Details of all the ORFs are shown in the Appendix \ref{ORFs}.

For the average response function $R_{II}(f)$ when $I = J$, a fully analytical expression of the Michelson-type combination has recently been obtained,
and the detailed derivations can be found in the
Refs.~\citep{Lu:2019log,Zhang:2020khm,Wang:2021jsv,Wang:2021owg,Wang:2022nea}.

%%%%%%%%%%%%%%%%%%%%%%%%%%%%%%%%%%%%%%%%%%%%%%%%%%%%%%%%%%%%%%%%%%%%%%%%%%
%%%%%%%%%%%%%%%%%%%%%%%%%%%%%%%%%%%%%%%%%%%%%%%%%%%%%%%%%%%%%%%%%%%%%%%%%%

\subsection{Likelihood}

%To facilitate detailed data analysis, In order to better evaluate the physical model parameters $\vec{\theta} \rightarrow \{\Omega_{\alpha}, \alpha, S^a, S^p\}$ later,
In the actual data analysis, we initially divide the observation data into a sequence of segments,
each with a duration of $\rm T_{seg}$ seconds and marked by the index $i$.
The strain data from the TianQin-LISA detector network, within the frequency domain, is represented as a vector $D_{i,k} \equiv D_{i}(f_{k}) = [\text{X}(f_{k}), \text{X}'(f_{k})]$,
where \( \text{X}(f_{k}) \) and \( \text{X}'(f_{k}) \) represent the strain components at frequency \( f_{k} \).
Accordingly, the likelihood function, which quantifies the probability of observing the data given certain parameters,
can be expressed as \citep{Romano:2016dpx, Adams:2010vc, Biscoveanu:2020gds}
\beq
\label{eq:lnlikeli}
{\cal L}_{i}(\vec{\theta}|D_{i})
=\prod_{k} \frac{1}{ \det|2\pi C_{i}(f_{k})|  }
\exp{\left\{-\frac{D_{i,k}  C^{-1}_{i}(f_{k}) D_{i,k}^{\dagger} }{2} \right\}}.
\eeq
Here, $\vec{\theta} \rightarrow  \{ \alpha, \Omega_{\alpha}, N_{a}, N_{p}\}$
represents the set of physical model parameters under consideration,
and $ D^{\dagger}$ denotes the conjugate transpose of the data vector.
%where $D = [X(f_{k}), X'(f_{k})]$ is the frequency-domain strain data stream from the two detectors,
%$^{\dagger}$ represents conjugate transpose,

The frequency- and $\vec{\theta}$-dependent covariance matrix, $C(f)$,
encompasses contributions from both the signal spectral density $S_{h}(f)$ and the instrumental noise spectral density $P_{n}(f)$:
%\begin{widetext}
\beqa
     C(f) =  \frac{{\rm T_{seg}}}{4}  \begin{pmatrix}
        \Gamma_{\rm XX} \, S_{ h} + P^{\rm tq}_{n}
        & \Gamma_{\rm XX'}\, S_{h}
        \\
        \Gamma_{\rm X'X}\, S_{h}
        & \Gamma_{\rm X'X'}\, S_{h}(f_{k})+P^{\rm lisa}_{n}
    \end{pmatrix}.
\eeqa
%\end{widetext}

When considering a set of $N$ data segments, the full likelihood ${\cal L}(\vec{\theta}|D)$ is derived as the product of the individual likelihoods associated with each segment,
\beq
\label{eq:totallikeli}
{\cal L}(\vec{\theta}|D) =\prod_{i}^{N} {\cal L}_{i}(\vec{\theta}|D_{i}).
\eeq

\section{Detection results }
\label{sec:Results}

In this section, we perform Bayesian inference on the \ac{GW} simulation data under three background signal models:
a power-law spectrum attributed to compact binary mergers,
and flat and single-peaked spectrum, both potentially originating from the early Universe.
Parameter estimation and detection limits are explored for these background signals.

For our simulation, we adopt the following assumptions to simplify the scenarios:
\begin{enumerate}
    \item[(1)] All resolvable sources, glitches, and other disturbances have been effectively subtracted from the raw data, leaving a data-set that includes only the \ac{SGWB} and instrumental noise.

    \item[(2)] The \ac{SGWB} and noise are assumed to be Gaussian, stationary, and uncorrelated in the frequency domain.
               Additionally, the instrumental noise parameters for both TianQin and LISA are symmetric, determined completely by the acceleration noise $S_{a}$ and the position noise $S_{p}$.
               %and are completely determined by acceleration noise $S^{a}(f)$ and position noise $S^{p}(f)$.
\end{enumerate}
And the injected values for the \ac{SGWB} and instrumental noise are listed in the second column of Table~\ref{tab:PL}.

\subsection{Parameter estimation}

In the Bayesian analysis framework, we assign a uniform prior to the spectral index and log-uniform priors to other parameters,
such as $\log_{10}\Omega_{\rm PL, Flat, SP}\in {\cal U}[-15, -9]$, $\log_{10} N_{a} \in {\cal U}[-43, -39]$, and $\log_{10} N_{p} \in {\cal U}[-53, -48]$ throughout this work.
For effective sampling of the high-dimensional posterior distribution, we employ the Markov chain Monte Carlo (MCMC) method,
specifically utilizing the affine-invariant ensemble sampler \texttt{emcee} \citep{Foreman-Mackey:2012any}.

%%%%%%%%%%%%%%%%%%%%%%%%%%%%%%%%%%%%%%%%%%%%%%%%%%%%%%%%%%%%%%%%%%%%%%%%%%%%%%
\begin{table}[htbp]%!htbp
    \centering
    \caption{The injection values and constraint results for the parameters of a SGWB and noise from TianQin and TianQin-LISA configurations.}
    \label{tab:PL}
    \begin{tabular}{c|c|c|c}
        \toprule
        Parameter  &\text{Injected} &\text{TianQin} &\text{TianQin+LISA} \\
        \midrule
        $n$   & $2/3 $ &$\quad\,\,0.642^{+0.209}_{-0.203}$   &$\quad\,\,0.678^{+0.048}_{-0.048}$ \\
        $\text{log}_{10} \Omega_{\rm PL}$       & $-11.357 $    &$-11.294^{+0.197}_{-0.209}$  &$-11.357^{+0.034}_{-0.033}$ \\
        $\text{log}_{10}N^{\text{tq}}_{a}$ &$-30.000$  &$-30.003^{+0.004}_{-0.004} $  &$-30.002^{+0.002}_{-0.002}$   \\
        $\text{log}_{10}N^{\text{tq}}_{p}$ & $-24.000 $ &$-24.000^{+0.001}_{-0.001}$  &$-24.000^{+0.001}_{-0.001}$ \\
        $\text{log}_{10}N^{\text{lisa}}_{a}$ & $-29.046 $ &$-$  &$-29.053^{+0.006}_{-0.006}$ \\
        $\text{log}_{10}N^{\text{lisa}}_{p}$ & $-21.648 $ &$-$  &$-21.648^{+0.001}_{-0.001}$ \\
        \hline
        $\text{log}_{10}\Omega_{\rm Flat}$ &$-11.000$  &$-10.932^{+0.056}_{-0.067}$  &$-10.993^{+0.012}_{-0.012}$ \\
        $\text{log}_{10}N^{\text{tq}}_{a}$ &$-30.000$  &$-30.003^{+0.004}_{-0.004}$  &$-30.001^{+0.002}_{-0.002}$ \\
        $\text{log}_{10}N^{\text{tq}}_{p}$ & $-24.000 $ &$-24.000^{+0.001}_{-0.001}$  &$-24.000^{+0.001}_{-0.001}$ \\
        $\text{log}_{10}N^{\text{lisa}}_{a}$ & $-29.046 $ &$-$  &$-29.051^{+0.006}_{-0.006}$ \\
        $\text{log}_{10}N^{\text{lisa}}_{p}$ & $-21.648 $ &$-$  &$-21.648^{+0.001}_{-0.001}$ \\
        \hline
        $\Delta$   & $\quad \, 0.200 $ &$\quad\,\,0.357^{+0.703}_{-0.131}$   &$\quad\,\,0.203^{+0.010}_{-0.009}$ \\
        $\text{log}_{10}\Omega_{\rm SP}$ &$-11.000$  &$-11.186^{+0.231}_{-1.014}$  &$-10.983^{+0.015}_{-0.015}$ \\
        $\text{log}_{10}N^{\text{tq}}_{a}$ &$-30.000$  &$-30.000^{+0.005}_{-0.005}$  &$-30.001^{+0.002}_{-0.002}$ \\
        $\text{log}_{10}N^{\text{tq}}_{\rm p}$ & $-24.000 $ &$-24.000^{+0.001}_{-0.001}$  &$-24.000^{+0.001}_{-0.001}$ \\
        $\text{log}_{10}N^{\text{lisa}}_{a}$ & $-29.046 $ &$-$  &$-29.050^{+0.005}_{-0.005}$ \\
        $\text{log}_{10}N^{\text{lisa}}_{\rm p}$ & $-21.648 $ &$-$  &$-21.648^{+0.001}_{-0.001}$ \\
        \bottomrule
    \end{tabular} %\\ \parbox[t]{2.0 \linewidth}{\flushleft\footnotesize ins. }
\end{table}
%%%%%%%%%%%%%%%%%%%%%%%%%%%%%%%%%%%%%%%%%%%%%%%%%%%%%%%%%%%%%%%%%%%%%%%%

\subsubsection{Astrophysical origin}

%%%%%%%%%%%%%%%%%%%%%%%%%%%%%%%%%%%%%%%%%%%%%%%%%%%%%%%%%%%%%%%%%%%%%%%%%
\begin{figure*}[htpb]
\centering
\includegraphics[width=0.9\textwidth,angle=0]{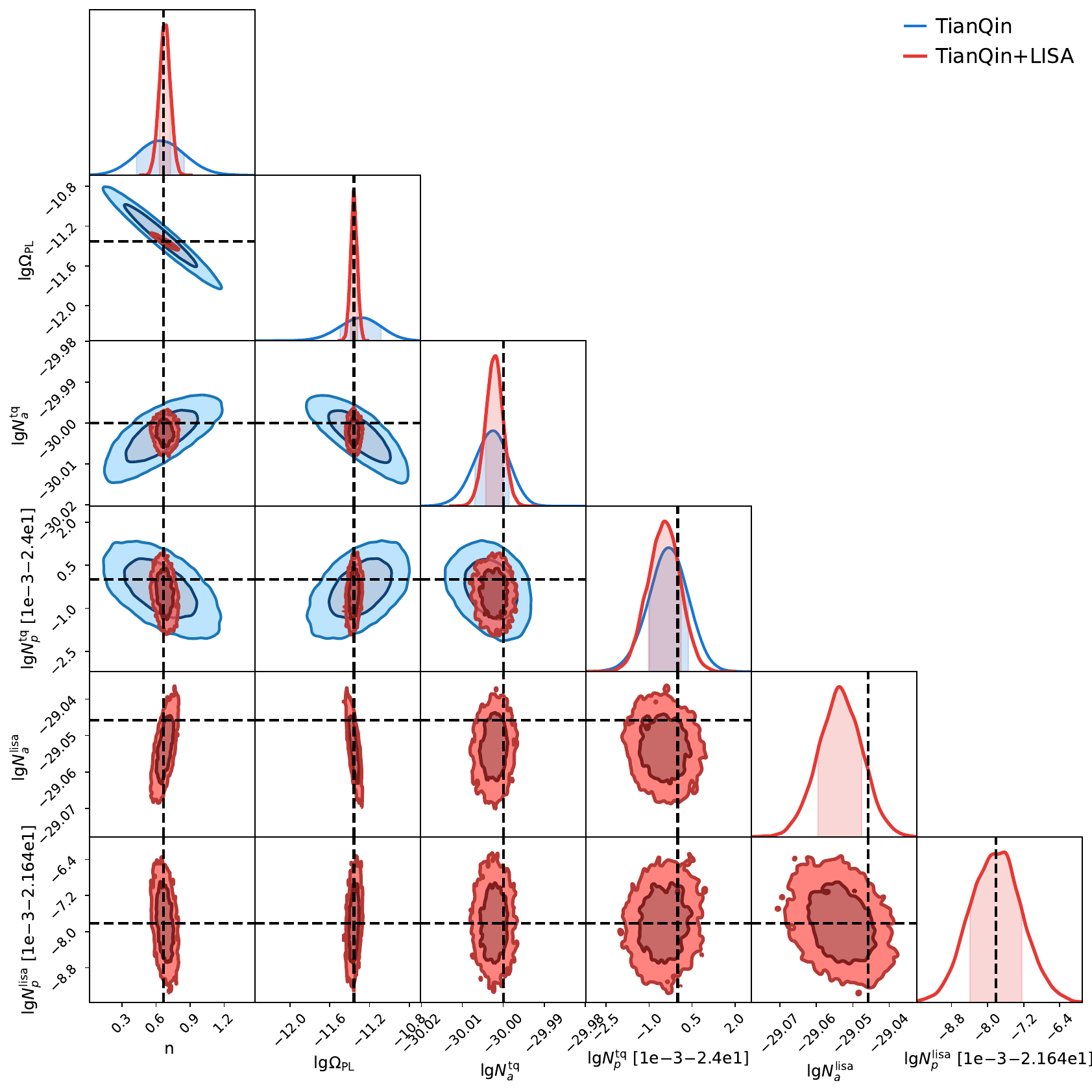}
\caption{Corner plot for the data = noise + a power-law spectrum case for TianQin-LISA detector network.
The black vertical dashed lines represent the injection values of the \ac{SGWB} and instrument noise parameters,
while the vertical shaded region on the posterior distribution denote 68\% credible region, the contour lines denote [68\%, 95\%] credible regions.}
\label{fig:SignalPL}
\end{figure*}
%%%%%%%%%%%%%%%%%%%%%%%%%%%%%%%%%%%%%%%%%%%%%%%%%%%%%%%%%%%%%%%%%%%%%%%%%

The \ac{CBC} are one of the most promising \ac{GW} sources for mHz space-based \ac{GW} observatories.
Analysis of numerous observations from current ground-based \ac{GW} detectors \citep{LIGOScientific:2018mvr,LIGOScientific:2020stg,LIGOScientific:2020zkf,LIGOScientific:2020iuh,LIGOScientific:2020aai}
reveals that \ac{CBC} could contribute to a detectable \ac{SGWB} for both LISA and TianQin \citep{Liang:2021bde}.
The analytic model describing the \ac{CBC} background signal depends on redshift and merger rates \citep{LIGOScientific:2017zlf,Regimbau:2011rp},
while the energy during the inspirals process can be characterised by a power-law (PL) spectrum \citep{Regimbau:2011rp,Farmer:2003pa,Moore:2014lga}, as detailed in Eqs.~(\ref{eq:PLspectrum}).
Here, we adopt fiducial values for the spectral index $n=2/3$ and the amplitude $\Omega_{\rm PL} = 4.4 \times 10^{-12}$ at the reference frequency $f_\mathrm{ref} = 1 \,\mathrm{mHz}$,
following previous studies \citep{LIGOScientific:2019vic,Chen:2018rzo}.

%for the parameters of the power-law spectrum background signal and instrument noise.
The corner plots presented in Fig.~\ref{fig:SignalPL} display the posterior distributions of the parameters for a power-law spectrum background signal and instrumental noise.
Within this figure, the injected values are denoted by black vertical dashed lines, while the 68\% and 95\% credible regions are outlined with contour lines.
The parameters  specifically the spectrum index $n$ and amplitude $\Omega_{\rm PL}$, are well constrained, predominantly falling within the 68\% credible interval, albeit with occasional deviations.

Combined with Table~\ref{tab:PL} and Fig.~\ref{fig:SignalPL}, it can be seen that the constraints on instrumental noise parameters are comparable for both the single TianQin detector and the TianQin-LISA detector network.
For the power-law \ac{SGWB}, the 68\% credible interval for both the spectrum index $n$ and amplitude $\Omega_{\rm PL}$ are larger than 0.2 in the single TianQin detector,
but they shrink to 0.05 and 0.03, respectively, in the TianQin-LISA detector network.
This indicates that incorporating the LISA detector would significantly improve the precision of the power-law \ac{SGWB} parameter estimates, with the credible intervals narrowing by a factor of 4 to 10.

\subsubsection{Cosmological origin}

Several mechanisms in the early Universe are believed to have contributed to the SGWB \citep{Maggiore:2000gv,Caprini:2018mtu},
including inflation \citep{Guth:1982ec,Bartolo:2016ami},
first-order phase transitions \citep{Hogan:1983ixn,Caprini:2015zlo,Caprini:2019egz},
and the dynamics of topological defect networks, such as cosmic strings \citep{Kibble:1976sj,Auclair:2019wcv}.
The cosmological \ac{SGWB} therefore serves as a valuable probe of the early Universe, with its detection promising to deepen our understanding of the fundamental physics shaping the early Universe and to provide insights into particle physics beyond the Standard Model \citep{ValbusaDallArmi:2020ifo}.

%%%%%%%%%%%%%%%%%%%%%%%%%%%%%%%%%%%%%%%%%%%%%%%%%%%%%%%%%%%%%%%%%%%%%%%%%
\begin{figure*}[htpb]
\centering
\includegraphics[width=0.9 \textwidth,angle=0]{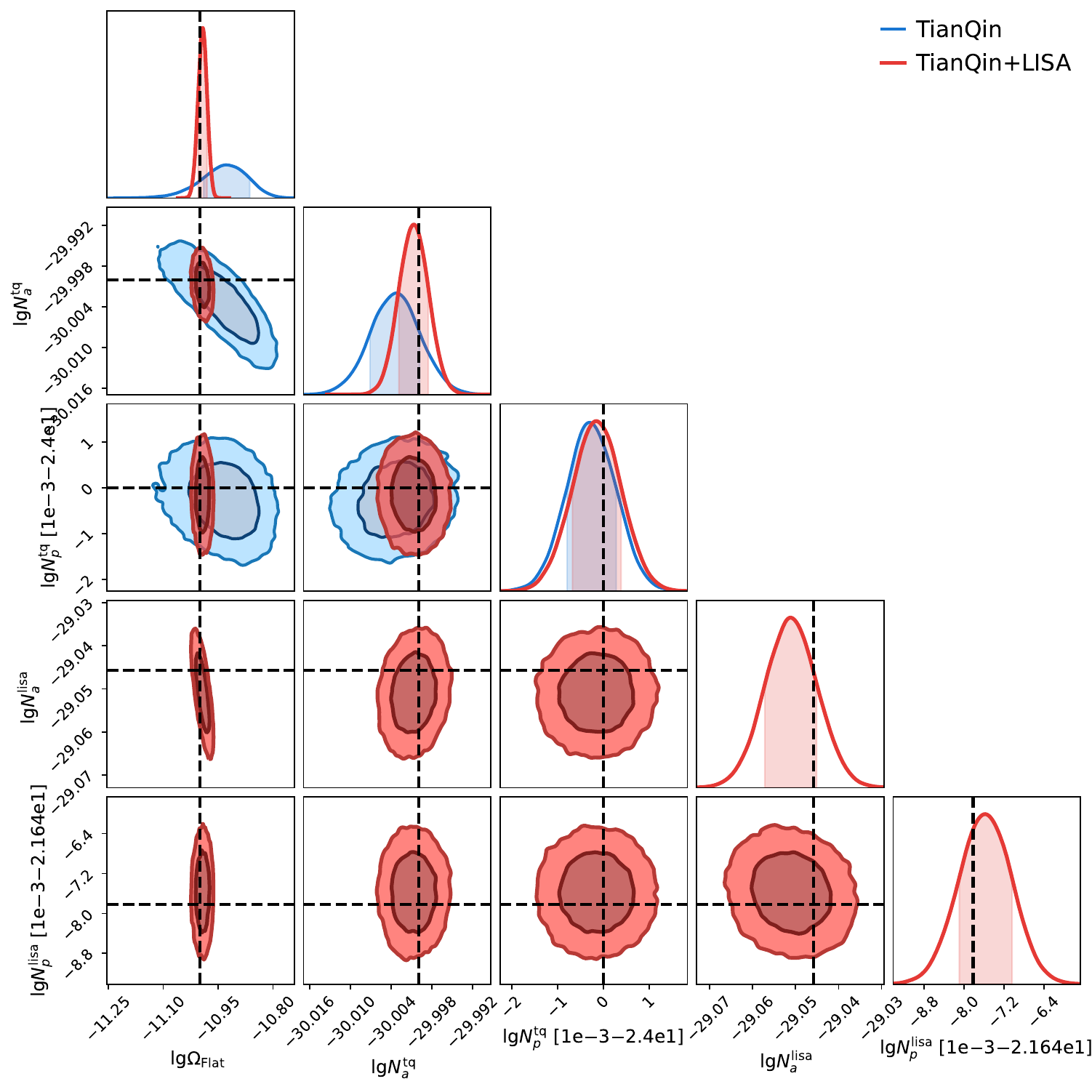}
\caption{Corner plot for the data = noise + a flat spectrum case for TianQin-LISA detector network.
The black vertical dashed lines represent the injection values of the \ac{SGWB} and instrument noise parameters.
In contrast, the vertically shaded region on the posterior distribution denotes 68\% credible region, and the contour lines denote [68\%, 95\%] credible regions.}
\label{fig:Signalflat}
\end{figure*}
%%%%%%%%%%%%%%%%%%%%%%%%%%%%%%%%%%%%%%%%%%%%%%%%%%%%%%%%%%%%%%%%%%%%%%%%%

%%%%%%%%%%%%%%%%%%%%%%%%%%%%%%%%%%%%%%%%%%%%%%%%%%%%%%%%%%%%%%%%%%%%%%%%%
\begin{figure*}[htpb]
\centering
\includegraphics[width=0.9\textwidth,angle=0]{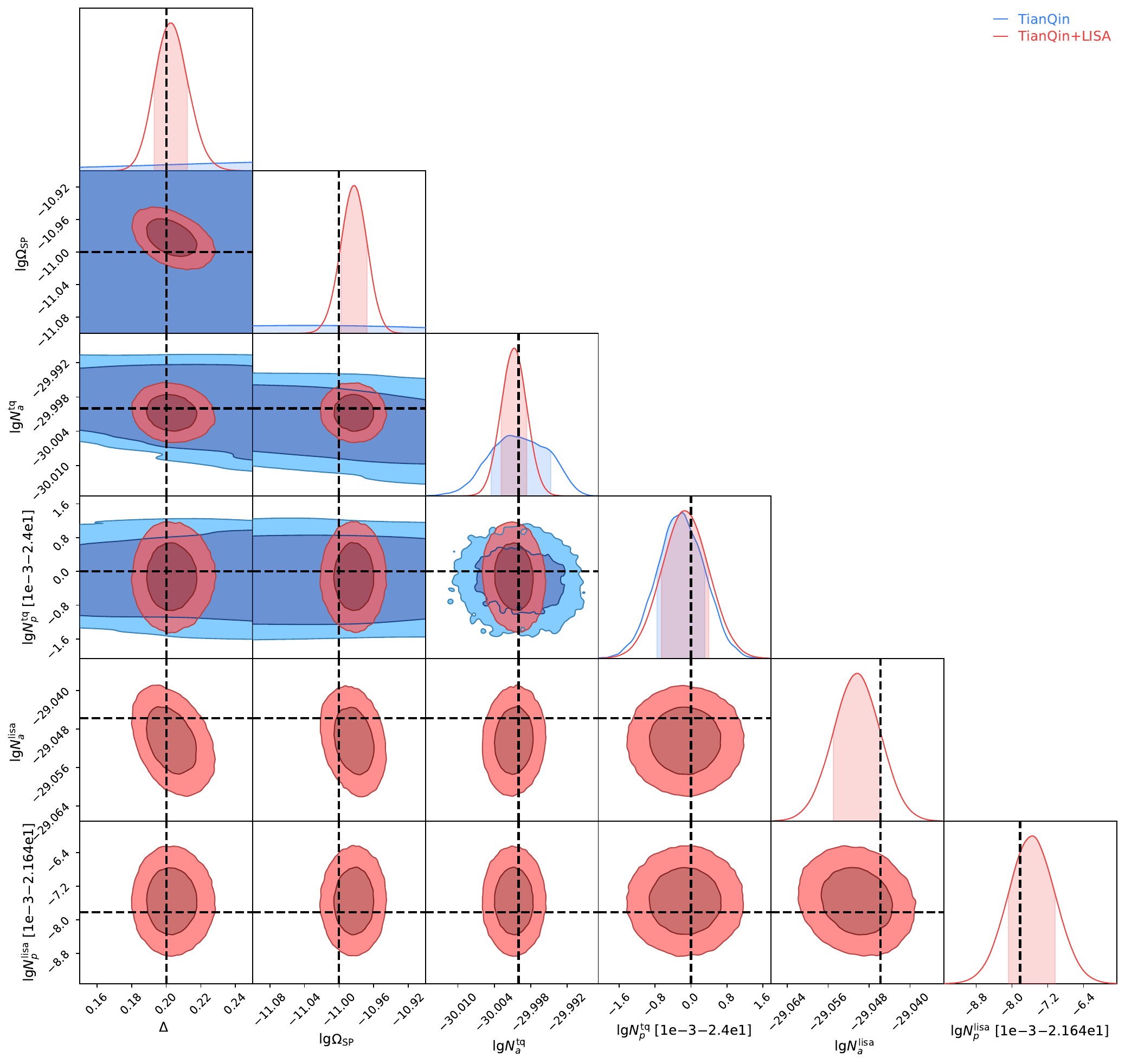}
\caption{Corner plot for the data = noise + a single-peak spectrum case for TianQin-LISA detector network.
The black vertical dashed lines represent the injection values of the \ac{SGWB} and instrument noise parameters. In contrast, the vertical shaded region on the posterior distribution denotes 68\% credible region, and the contour lines denote [68\%, 95\%] credible regions.}
\label{fig:SignalSP}
\end{figure*}
%%%%%%%%%%%%%%%%%%%%%%%%%%%%%%%%%%%%%%%%%%%%%%%%%%%%%%%%%%%%%%%%%%%%%%%%%

In the TianQin-LISA networks detection band, a scale-invariant (flat spectrum) cosmological \ac{SGWB} can be generated by
the amplification of vacuum fluctuations during slow-roll inflation \citep{Grishchuk:1974ny,Grishchuk:1993ds,Starobinsky:1979ty,Maggiore:1999vm}.
The amplitude of these signals depends on the primordial fluctuation power spectrum formed in the early Universe.
In this paper, we choose $\Omega_{\rm Flat} = 1.0 \times 10^{-11} $ as the fiducial value. %, as detailed in Eqs.~(\ref{eq:SPspectrum})
In addition, for the Gaussian-bump spectrum SGWB, we adopt
$\Omega_{\text{SP}} = 1.0 \times 10^{-11}$, $\Delta = 0.2$,
$f_{\text{ref}} = 3$ mHz as the inject parameters \citep{Cheng:2022vct,Caprini:2019pxz,Flauger:2020qyi}.

The corner plot for the scenario in which the data consists of a cosmological origin  \ac{SGWB} are shown in Fig. \ref{fig:Signalflat} and Fig. \ref{fig:SignalSP},
and the constraints on the parameters, derived from different detector configurations, are detailed in Table~\ref{tab:PL}.
Similar to the power-law scenario, all parameters are well constrained within a 1 $\sigma$ confidence interval for the TianQin-LISA detector.
In addition, the constraints on the instrumental noise parameters are still comparable between the TianQin and the joint network.
However, the credible interval for the amplitude parameter $\log_{10}\Omega_{\rm Flat}$ is reduced by a factor of 5 within the TianQin-LISA detector network.
Moreover, by comparing table \ref{tab:PL}, we find that the flat background signal here has a higher parameter estimation accuracy,
meaning that it is better distinguished from instrument noise than the injected power-law spectrum.

The single-peak SGWB scenario, illustrated in Fig. \ref{fig:SignalSP} and table \ref{tab:PL}, is the most distinctive compared to other models considered in this study.
With individual TianQin, the amplitude parameters, $\log_{10}\Omega_{\rm SP}$ and $\Delta$, characterizing the single-peak SGWB, are poorly constrained.
The inclusion of the LISA detector, however, significantly improves the constraints on these parameters.
Assuming a three-month observation, the TianQin-LISA detector network is expected to confidently detect
a single-peak SGWB with an energy density as low as $\Omega_{\rm SP} = 1.2 \times 10^{-12}$,
as demonstrated in the right plane of Fig. \ref{fig:BayesFactor}.

\subsection{Detection limit}

%%%%%%%%%%%%%%%%%%%%%%%%%%%%%%%%%%%%%%%%%%%%%%%%%%%%%%%%%%%%%%%%%%%%%%%%%%
\begin{figure*}[htbp]%[!ht]
\centering
    \includegraphics[width=0.32\linewidth,trim={0.4cm 0 1.2cm 1.1cm}, clip]%
    {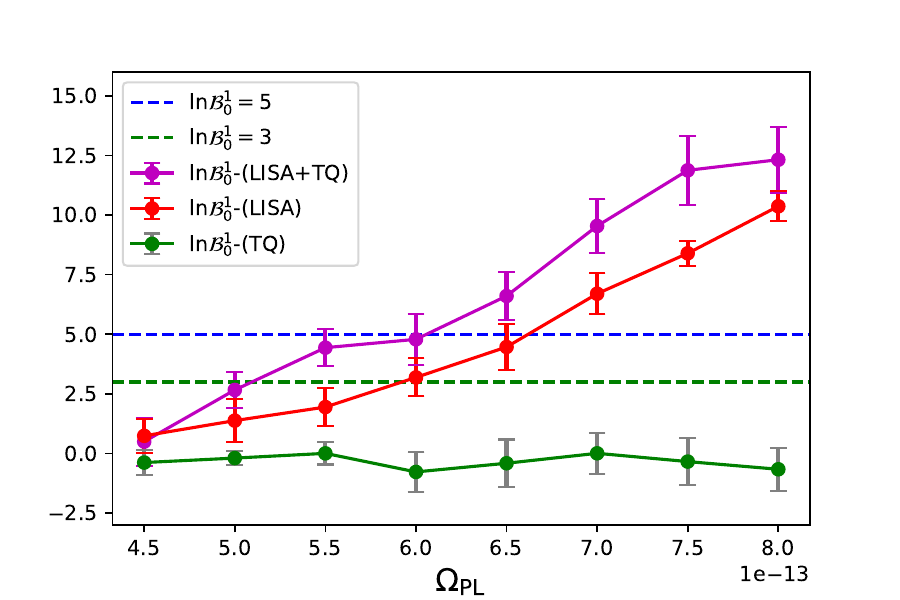}
    %{PLBayesFactor.pdf}
    \includegraphics[width=0.32\linewidth,trim={0.4cm 0 1.2cm 1.1cm}, clip]%
    {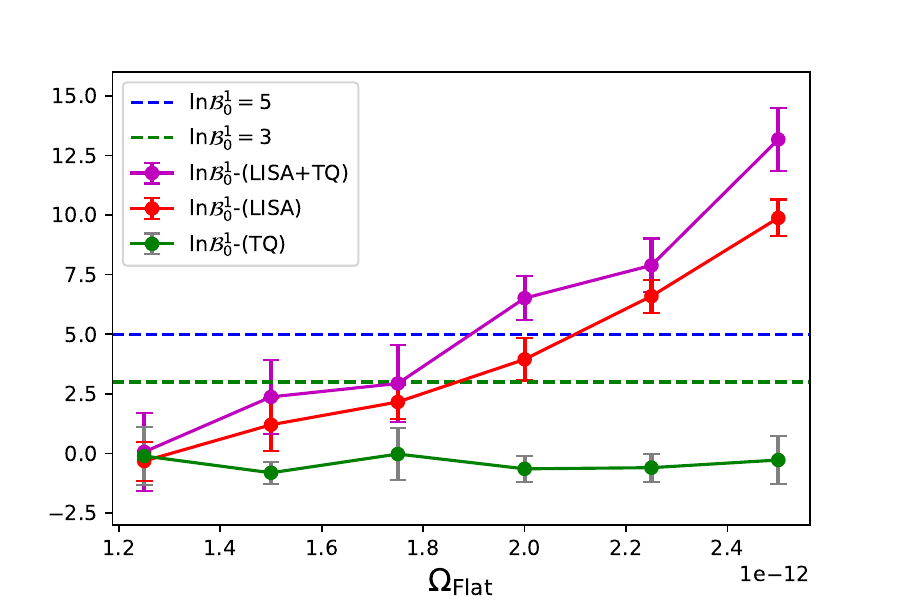}
    %{FlatBayesFactor.pdf}
    \includegraphics[width=0.32\linewidth,trim={0.4cm 0 1.2cm 1.1cm}, clip]%
    {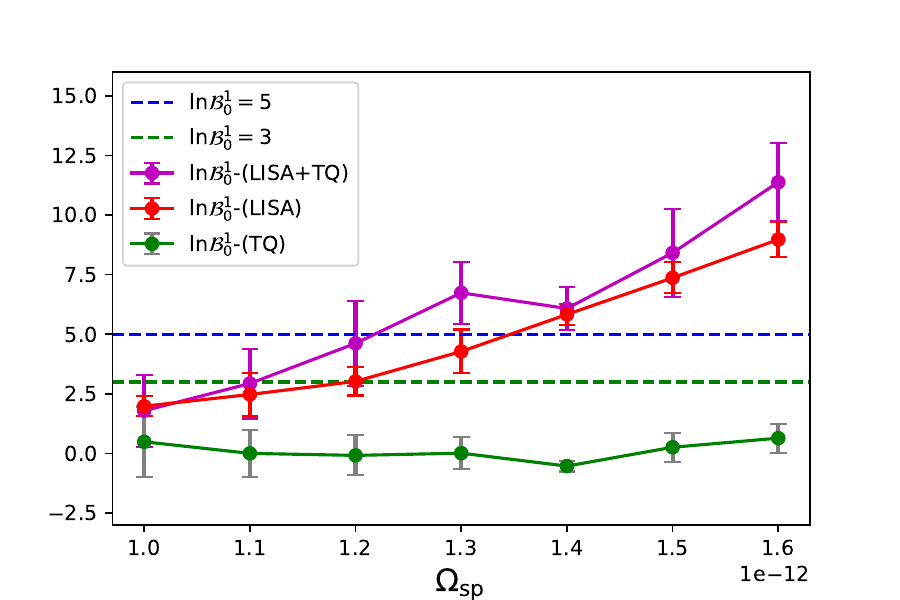}
    %{SPBayesFactor.pdf}
   \caption{The Bayes factor as a function of background amplitude in the Michelson-X channel,
    which showing the detectability versus level of a power law background (left plane), a flat background (center plan) and a sing-peak background(right plane).
    The green and blue dotted lines represent the log-Bayes factors 3 and 5, respectively. The magenta line represents the Bayes factor.}
\label{fig:BayesFactor}
\end{figure*}
%%%%%%%%%%%%%%%%%%%%%%%%%%%%%%%%%%%%%%%%%%%%%%%%%%%%%%%%%%%%%%%%%%%%%%%%%%

The exploration of the detection capabilities of the TianQin-LISA detector network for certain theoretical models involves not only parameter estimation,
but also challenges related to detection limits and model selection, i.e. discerning which model is more strongly supported by the actual observational data.
%which model best matches the observed data.

This is usually done by calculating the \emph{Bayes factor} $\mathcal{B}^{1}_{0}$, which is the ratio of the posterior probabilities of the models $\mathcal{M}_1$ and $\mathcal{M}_0$ given the observed data $D$:
\begin{align*}
\mathcal{B}^{1}_{0} &= \frac{p(\mathcal{M}_1|D)}{p(\mathcal{M}_0|D)} = \frac{p(D|{\cal M}_1)}{p({D|\cal M}_0)}  \frac{p({\cal M}_1)}{p({\cal M}_0)},
\end{align*}
where
\begin{itemize}
    \item $\mathcal{M}_{0}$: ~the observed data is attributed to instrument noise only,
    \item $\mathcal{M}_{1}$: the observed data includes both instrument noise and a \ac{SGWB}.
\end{itemize}

To compare the two models, the prior odds  $ \frac{p({\cal M}_1)}{p({\cal M}_0)}$ is set to unity, and thus we focus on the Bayes factor.
The Bayes factor is calculated by evaluating the evidence under both models using the \ac{NS} algorithm \texttt{dynesty}, as detailed in \citep{2020MNRAS.493.3132S}.

A positive log Bayes factor $\log {\cal B}^{1}_{0}$ shows support for the ${\cal M}_1$ over ${\cal M}_0$.
But to avoid the influence of random fluctuation, it is widely suggested that a value of $\log {\cal B}^{1}_{0}>1$  is required for a meaningful follow-up discussion, and a value of $\log {\cal B}^{1}_{0}>3$ is needed for strong support of the ${\cal M}_1$ \citep{Kass:1995loi}.

Fig. \ref{fig:BayesFactor} presents the variation of the log-Bayes factor as a function of background amplitude in the Michelson-X channel.
Three spectral shapes are considered: a power-law background (left panel), a flat background (center panel), and a single-peak background (right panel).
In each panel, horizontal dashed lines denote the log-Bayes factors of 3 and 5, respectively, serving as benchmarks for evaluating the statistical significance of the observations.
To ensure a statistically consistent comparison, the Bayes factor for TianQin (TQ, green), LISA (red), and the combined TianQin–LISA network (TQ+LISA, magenta) has been evaluated with the same pseudo-random seed and identical injected amplitude, assuming a three-month observation period.
Except for TianQin, the Bayes factor increases with the background amplitude in all three panels, indicating enhanced detectability as the background becomes larger.
The comparison across different configurations (TQ+LISA, LISA, TQ) reveals that the Bayes factor is generally higher for the combined TQ+LISA network, suggesting better detectability when data from both TianQin and LISA are considered together.
Specifically, the Bayesian factor of the TQ+LISA network is slightly higher than that of LISA and significantly higher than that of TianQin. This indicates that the joint exploration will bring a substantial enhancement to TianQin's capabilities, highlighting the potential benefits of combining observations from multiple detectors.
Our detection confidence becomes very strong (a log-Bayes factor of 5) for a power-law background level $\Omega_{\rm PL} = 6.0 \times 10^{-13}$,
a flat background level of $\Omega_{\rm Flat} = 2.0 \times 10^{-12}$,
and a single-peak background level $\Omega_{\rm SP} = 1.2 \times 10^{-12}$
with three-month data  (shown in Fig. \ref{fig:BayesFactor}).

\section{Summary and discussion}
\label{sec:SUMMARY AND DISCUSSION}

In this paper, we developed a Bayesian data analysis framework for \ac{SGWB} search based on the Michelson-X channels of two space-based \ac{GW} detector,
and take future TianQin-LISA detector network as an example to study its detectability.
We employ a numerical method to calculate the overlap reduction functions and assess the detectability of the TianQin-LISA detector network for isotropic \ac{SGWB}
compared to that of a single TianQin.
To perform these investigations, three models for \ac{SGWB} spectrum are considered:
a power-law spectrum arising from compact binary coalescences,
a scale-invariant flat spectrum characteristic of inflationary models,
and a single-peak spectrum.
Using MCMC sampling, the noise and \ac{SGWB} parameters were precisely
recovered by both the TianQin-LISA detector network and the single TianQin detector.
Our results demonstrate that the inclusion of the LISA would contribute to improving the \ac{SGWB}
detection sensitivity and parameter estimation accuracy compared to the single TianQin detector.

To assess the detection limit of the cross-correlation method for detecting the \ac{SGWB},
we apply Bayesian model selection to compare a model consisting of noise plus a \ac{SGWB} with one containing noise only.
We calculate the Bayes factors across a range of \ac{SGWB} amplitudes,
and concluded that the TianQin-LISA detector network will be capable of detecting the energy density of
a power-law spectrum signal down to $\Omega_{\rm PL} = 6.0 \times 10^{-13}$,
a flat signal down to $\Omega_{\rm Flat} = 2.0 \times 10^{-12}$,
and a single-peak signal down to $\Omega_{\rm SP} = 1.2 \times 10^{-12}$,
all within a three-month observational period.

The current discussion focuses on distinguishing single-component \ac{SGWB} from instrument noise and does not yet address the simultaneous identification of astrophysical and cosmological sources.
Our future research will extend in two directions: first, to the optimal \ac{TDI} combinations of the \{A, E, T\} channels,
derived from the three Michelson-type \{X, Y, Z\} channels;
and second, to investigate how these limits are affected by astrophysical confusion foregrounds. We will leave these more realistic scenarios for future work.

%%==================================
%%\acknowledgments
\section*{Acknowledgements}

This work has been supported by the National Key Research and Development Program of China (No. 2023 YFC 2206700),
and Doctoral Research Start-up Project of Hunan University of Arts and Science (No. 23BSQD26).

%%@@@@@@@@@@@@@@@@@@@@@@@@@@@@@@@@@@@@@@@@@
\appendix
\section{Time-dependent ORFs in Michelson-X channel}
%The ORFs in Michelson X channel
\label{ORFs}

\begin{figure*}[!htbp]%[!ht]
    \includegraphics[width=7.5 cm,angle=0,clip]{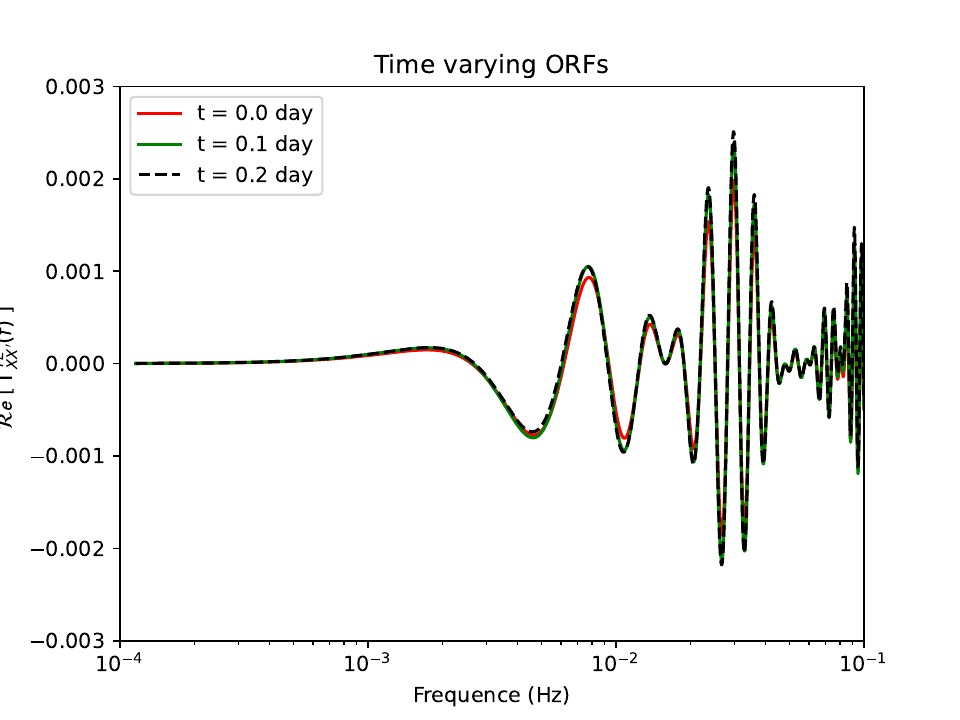}
    \includegraphics[width=7.5 cm,angle=0,clip]{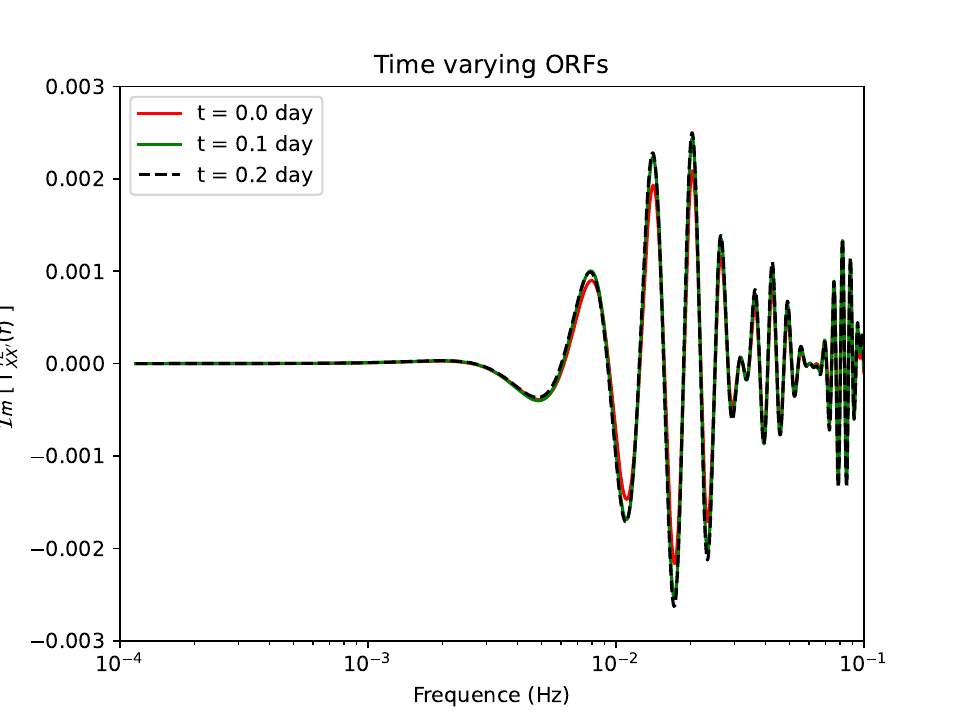}
    \\
    \includegraphics[width=7.5 cm,angle=0,clip]{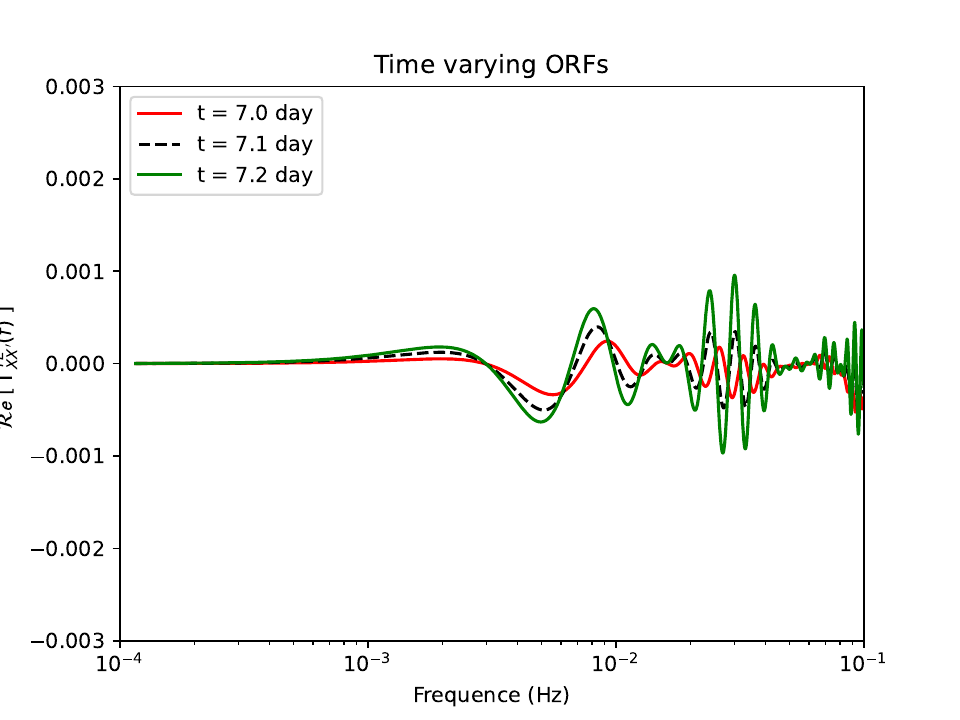}
    \includegraphics[width=7.5 cm,angle=0,clip]{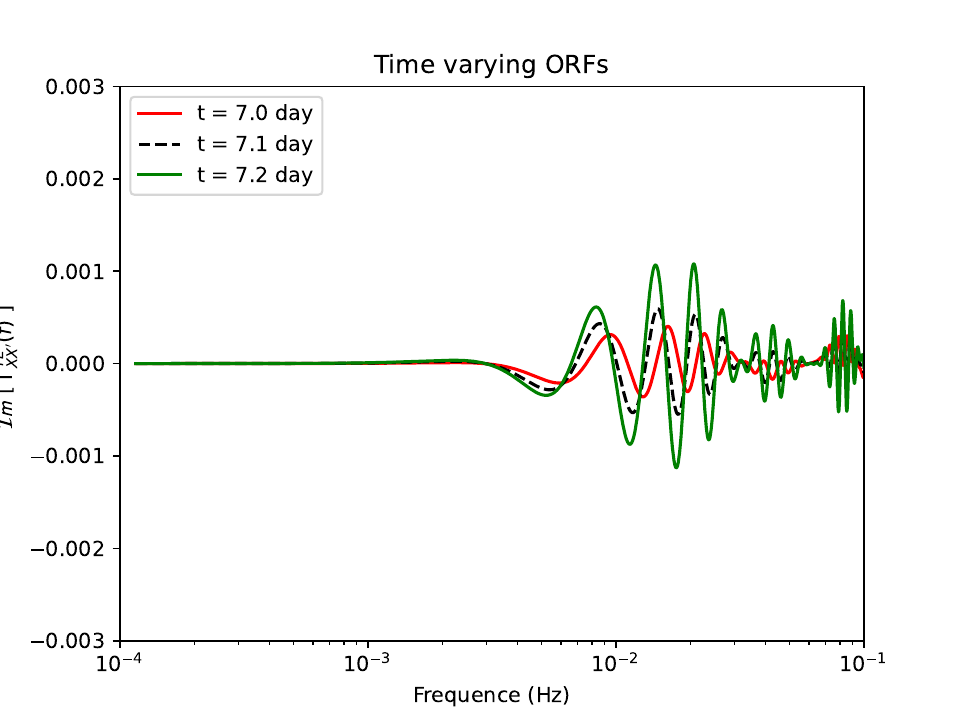}
   \caption{The overlap reduction function as a function of time in X channel,
    which shows the detector network's correlated responses to the \ac{SGWB} at t = 7.0 day, t = 7.1 day, and t = 7.2 day.
    The left panel illustrates the real parts, while the right panel represents the imaginary parts.}
\label{fig:ReIMORFs}
\end{figure*}

\begin{figure*}[!htbp]%[!ht]
    \includegraphics[width=7.5 cm,angle=0,clip]{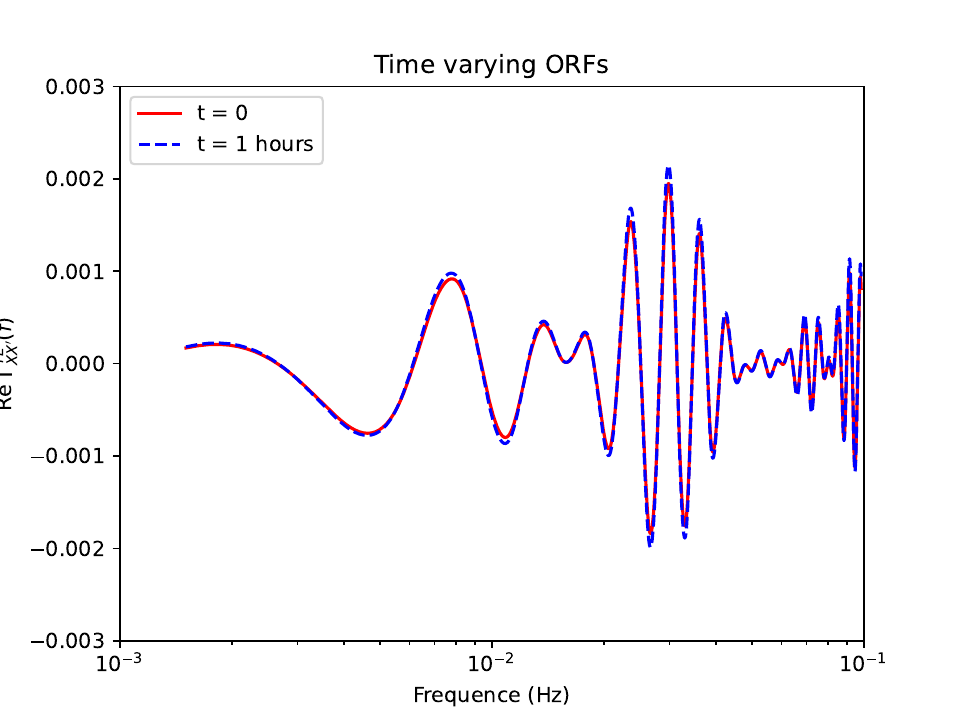}
    \includegraphics[width=7.5 cm,angle=0,clip]{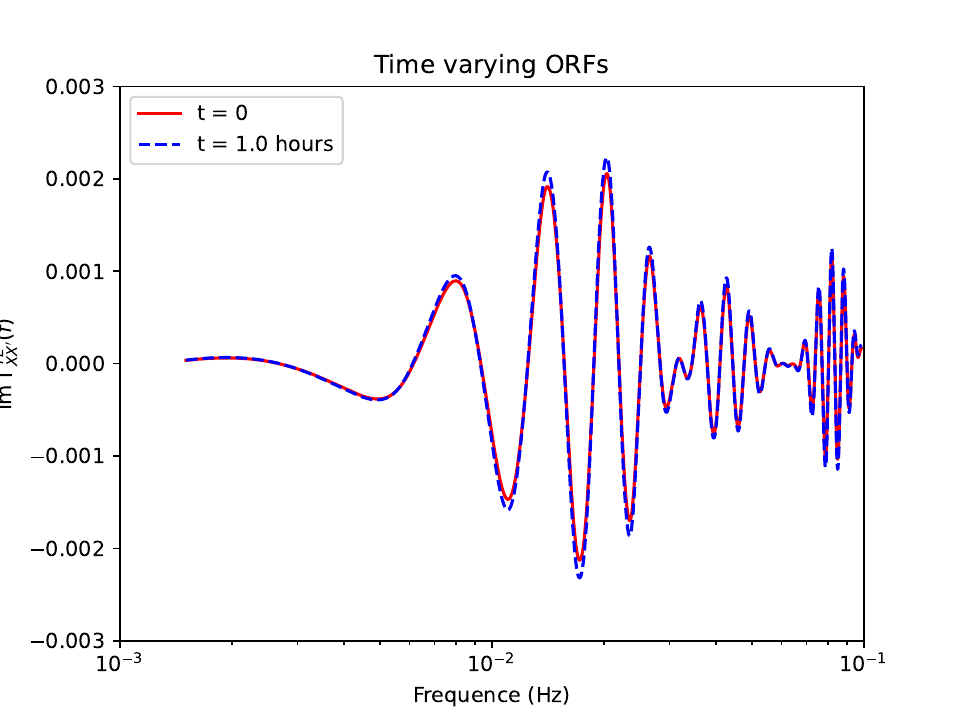}
   \caption{The overlap reduction function as a function of time in X channel,
    which shows the detector network's correlated responses to the \ac{SGWB} at t = 0, and t = 1 hours.
    The left panel illustrates the real parts, while the right panel represents the imaginary parts.}
\label{fig:ReIMORFhours}
\end{figure*}

In space-based \ac{GW} detector networks, the relative positions of detectors change over time, so that the ORFs will also be time-dependent.
This characteristic introduces new challenges for detecting the \ac{SGWB}.
To tackle this problem, we segment the total observational data and analyze the time-varying ORFs across different intervals (shown in Fig. \ref{fig:ReIMORFs} and \ref{fig:ReIMORFhours}).

%%%%%%%%%%%%%%%%%%%%%%%%%%%%%%%%%%%%%%%%%%%%%%%%%%%%%%%%%%%%%%%%%%%%%%%%%%
%%%%%%%%%%%%%%%%%%%%%%%%%%%%%%%%%%%%%%%%%%%%%%%%%%%%%%%%%%%%%%%%%%%%%%%%%%

%%%%%%%%%%%%%%%%%%%%%%%%%%%%%%%%%%%%%%%%%%%%%%%%%%%%%%%%%%%%%%%%%%%%%%%%%%
%%%%%%%%%%%%%%%%%%%%%%%%%%%%%%%%%%%%%%%%%%%%%%%%%%%%%%%%%%%%%%%%%%%%%%%%%%

\bibliography{TL-SGWB}

\end{document}